\magnification 1200
\centerline {REEXAMINATION OF SEVEN-DIMENSIONAL SITE PERCOLATION THRESHOLDS}

\bigskip
Dietrich Stauffer$^1$
and
Robert M. Ziff$^2$
\bigskip
$^1$ Institute for Theoretical Physics, Cologne University, D-50923 K\"oln,
Euroland

$^2$ Department of Chemical Engineering, University of Michigan, Ann Arbor,
MI 48109-2136, USA

e-mail: stauffer@thp.uni-koeln.de, rziff@umich.edu

\bigskip
Monte Carlo simulations alone could not clarify the corrections to scaling for
the size-dependent $p_c(L)$ above the upper critical dimension. Including
the previous series estimate for the bulk threshold $p_c(\infty)$ gives
preference for the complicated corrections predicted by renormalization group
and against the simple $1/L$ extrapolation. Additional Monte-Carlo simulations
using the Leath method corroborate the series result for $p_c$.
\noindent

Keywords: Monte Carlo, 1/d expansion, finite-size scaling

\bigskip
How to extrapolate from finite samples of linear dimension $L$ to the
thermodynamic limit $L \rightarrow \infty$ is an important problem in physics
simulations. Above the upper critical dimension (4 in usual Ising models, 6 in
random percolation), the critical exponents are known but nevertheless
controversies remain$^1$ for the five-dimensional Ising model. Earlier, for site
percolation in seven dimensions, the numerical variation of the
apparent threshold $p_c(L) = \langle p_c \rangle$
on a hypercubic lattice with $L^7$ sites could be fitted$^2$ on
$p_c(\infty)-p_c(L) \propto 1/L$, but also$^3$ on the theoretical prediction
$p_c(\infty)-p_c(L) \propto 1/L^2 - {\rm const}/L^{7/3}$. Also for
three-dimensional self-avoiding walks
one may assume one empirical correction term, or many terms as predicted by
renormalization methods with more free parameters$^4$. Therefore the present
note reexamines seven-dimensional site percolation in the search of a
clarifying example, using a much better Cray-T3E computer and a slightly
improved program$^5$ compared with ref.2.

We checked if the top of the hypercube is connected with the bottom hyperplane,
using helical boundary conditions in five and free boundaries in the two
remaining directions$^5$ and a Hoshen-Kopelman algorithm without recycling$^6$.
Random number generation by integer multiplication
with 16807 gave problems for $L=4, \, 8,\, 16$ even for 64-bit arithmetic, but
otherwise agreed with results from the Kirkpatrick-Stoll R250 generator; only
the latter method was used for the data shown here. About 640 samples were
averaged over for each $L$, giving an accuracy of about $10^{-4}$ for $p_c$ in
large lattices. ($[<p_c^2> - <p_c>^2]^{1/2}$ was not investigated in
detail since it seems to vary as $1/L^2$.) Fig.1 shows our results for $L$ up
to 20, together with a linear fit
$$ p_c(L) = 0.0909 - 0.12/L \eqno (1)$$
and {\it the same} fit as published before$^3$:
$$ p_c(L) = 0.0887 - 3.3/L^2 + 5/L^{7/3} \quad .\eqno (2)$$

We see that for large $L$ both fits agree nicely with our new Monte Carlo
results: On the basis of these simulations alone no preference for one fit
over the other is visible. However, when we extract from series expansions$^7$
(in terms of reciprocal dimensionality) a bulk $p_c$ of about 0.089, then the
more complicated fit (2) is clearly preferred over the linear fit (1). This
is a nice example how the combination of series with Monte Carlo techniques
can give some answers which are not clear from each technique separately$^8$.
However, a more direct and new series expansion determination of $p_c$ would
be desirable.

If we would have simulated $L = 40$ instead of $L = 20$ we might have chosen
between the two fits on the basis of simulations alone. Such a size would
have required about two orders of magnitude more in computer time and memory
than the about 4000 processor hours (512 Mbyte each) used here.

(About boundary conditions: If we check whether top and bottom are connected
we cannot use vertical periodic boundary conditions. As long as not all
directions are treated with periodic boundary conditions, and at least one
direction uses free boundaries, the finite-size effects are expected to be
dominated by the
free boundaries. The bulk $p_c$ value is independent of the boundary
conditions, and so may be the powers of $L$ in finite-size corrections,
but finite-size amplitudes as well as the fraction of samples spanning at $p_c$ 
depend on such details.$^9$ The Hoshen-Kopelman algorithm stores only one line 
of a square lattice at any one moment, and then it is simplest to use free 
boundaries also in this direction. In higher dimensions it is practical to
store the $L^{d-1}$ sites of the one hyperplane kept in memory by a 
one-dimensional index $i=1,2, \dots, L^{d-1}$. In this way $^5$ we come to
our mixture of helical and free boundary conditions.) 
  
Thus, to show that Monte Carlo can compete with series expansions, we made
a different determination of the bulk $p_c$ using the Leath cluster growth
algorithm$^{10}$. Here, when a cluster stops growing before touching the
boundaries, its properties are completely free of finite-size effects,
and such data cannot be directly compared with the above Hoshen-Kopelman
data for $p_c(L)$. However, the bulk $p_c$ must be the same.
We determined $p_c$ by the methods given recently for three-dimensional
lattices$^{11}$.  We considered a virtual lattice of size $32^7$
and a maximum size cut-off of 16384 sites, and simulated 
$10^6$ -- $10^7$ clusters each at various values of $p$.
The clusters are grown independently for each value of $p$,
since in this method the random numbers are not assigned
to all lattice sites but only to those the cluster visits. 

It turned out that some of the largest clusters clusters (of maximum
size of 16384 sites) wrapped a bit around the periodic boundary of
the system, which could conceivably lead to the cluster touching
itself and cause a bias.  However, because of the very large number
of sites in the lattice, the probability of this occurring is very
low.  (We could not easily check for wraparound error in our program.)
In any case, the data for smaller clusters (where no wraparound was
even possible) and for $s=16485$ was completely consistent.

Near the critical point, $P_{\ge s}$, defined as the probability
that a cluster grows to a size greater than or equal to $s$, behaves as
$$ P_{\ge s} \sim s^{2-\tau} f((p-p_c)s^\sigma) \approx
s^{2-\tau} (A + B (p-p_c)s^\sigma +\ldots)  \eqno(3) $$
where $\tau = 5/2$ and $\sigma = 1/2$ for $d \ge 6$. \  Thus in Fig.~2 we
plot $s^{1/2}P_{\ge s}$ vs.\ $s^{1/2}$ for $p = 0.0885,$ 0.0888, 0.0889 and 
0.0890, and find good agreement with the expected behavior with $A \approx 
1.44$, $B \approx 20$, and
$$ p_c = 0.08893 \pm 0.00002 \eqno (4)$$
which is in  excellent agreement with the series extrapolation$^7$.
With this value,
the simple fit of eq (1) can be excluded since it requires $p_c = 0.091$.

\bigskip

Acknowledgements

\noindent
We thank A. Aharony, D.L. Hunter and N. Jan for discussions leading to this
work, which was supported by GIF and the German Supercomputer Center, J\"ulich.
RZ acknowledges support from the U. S. National Science Foundation under grant
DMR-9502700.

\bigskip
\parindent 0pt
References:

1.
H.W.J.Bl\"ote and E. Luijten, Phys. Rev. Lett. 76, 3662 (1996) and
Europhys. Lett.  38, 565 (1997);
X.S. Chen and V. Dohm, Int. J. Mod. Phys. C 9, 1007 and 1073 (1998);
E. Luijten, K. Binder and H. W. J. Bl\"ote, Eur. Phys. J. B 9, 289 (1999);
M. Cheon, I. Chang and D. Stauffer, Int. J. Mod. Phys. C 10, 131 (1999).

2. D. Stauffer,  Physica A 210, 317 (1994).

3. A. Aharony and D. Stauffer, Physica A 215, 342 (1995).

4. S. Joseph, D. L. Hunter, D. MacDonald, L.L. Moseley,
N. Jan and T. Guttmann, preprint

5. D. Stauffer and N. Jan, in: {\it Annual Reviews of Computational Physics},
vol. VIII (Zanjan proceedings), World Scientific, Singapore 2000.

6. D. Stauffer and A. Aharony, {\it Introduction to Percolation Theory},
Taylor and Francis, London 1994; A. Bunde and S. Havlin, {\it Fractals and
Disordered Systems}, Springer, Berlin-Heidelberg 1996; M. Sahimi, {\it
Applications of Percolation Theory}, Taylor and Francis, London, 1994.
For recent percolation thresholds see S.C. van der Marck, Int. J. Mod. Phys. C
9, 529 (1998).

7. D.S. Gaunt, M.F. Sykes and H.J. Ruskin, J. Phys. A 9, 1899 (1976).

8. J. Adler, page 241 in: {\it Annual Reviews of Computational Physics},
vol. IV, World Scientific, Singapore 1996.

9. M. Ford, D.L. Hunter, and N. Jan, Int. J. Mod. Phys. C 10, 183 (1999).

10. P.L. Leath, Phys. Rev. B 14, 5046 (1976).

11. C.D. Lorenz and R.M. Ziff, Phys. Rev. E 57, 230 (1998).

\bigskip

Fig.1: Effective thresholds $p_c(L)$ versus $1/L$, together with eqs.(1,2). Part
a gives the overall trend, part b expands the region of large lattices. The
horizontal line is the series extrapolation for infinite $L$.

Fig.2: Plots of $s^{1/2}P_{\ge s}$ vs $s^{1/2}$ from the Leath
simulations, for  $p = 0.0890,$ 0.0889, 0.0888, and 0.0885 from top to bottom.
The slopes of the three upper curves are 0.0014, -0.0006 and -0.0026 
respectively.

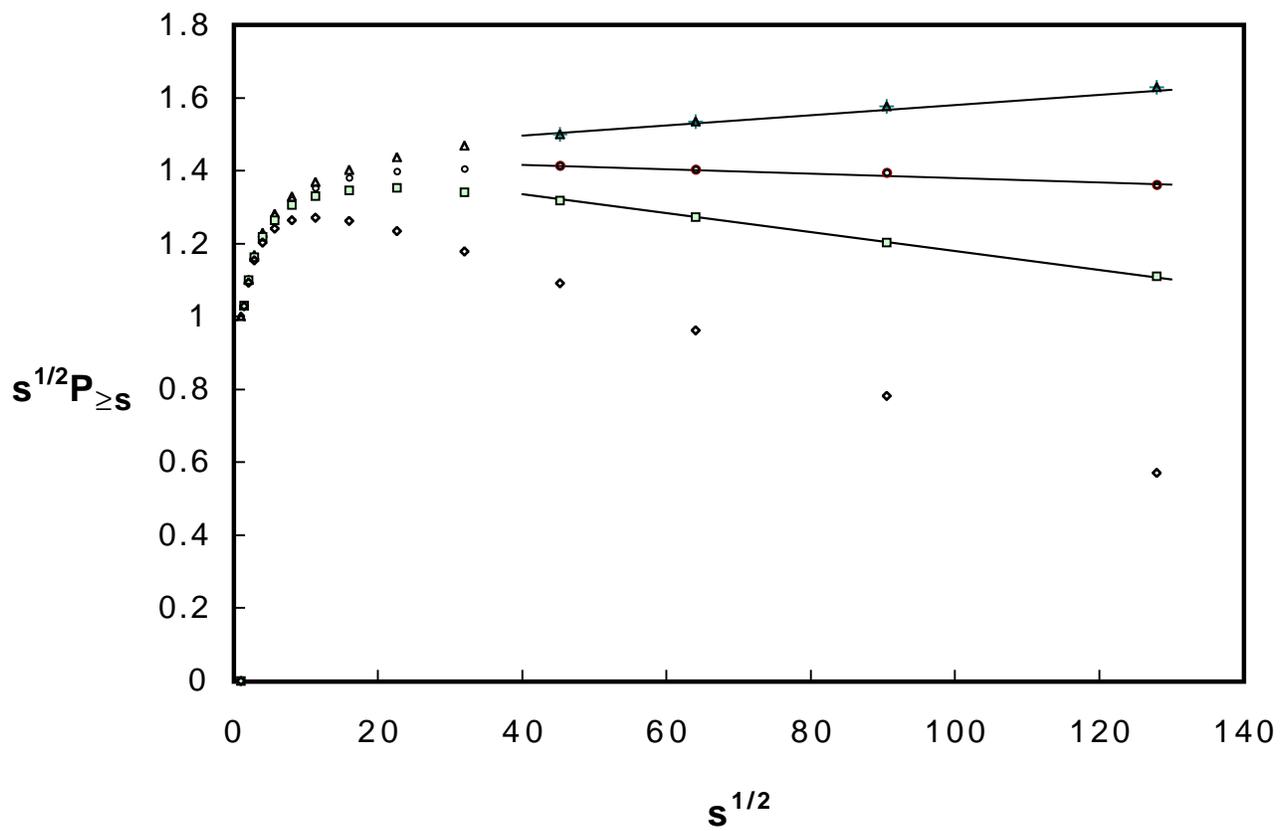
\end